\documentclass[a4paper,12pt]{amsart}

\usepackage{amsmath}
\usepackage{amsfonts}
\usepackage{amssymb}
\usepackage{amsthm}
\usepackage{tikz}
\usepackage{float}

\newtheorem{theorem}{Theorem}[section]

\newtheorem{corollary}[theorem]{Corollary}
\newtheorem{lemma}[theorem]{Lemma}

\theoremstyle{definition}
\newtheorem{definition}[theorem]{Definition}

\theoremstyle{remark}
\newtheorem{remark}[theorem]{Remark}
\newtheorem{example}[theorem]{Example}

\newcommand{\p}{\partial}
\newcommand{\la}{\left\langle}
\newcommand{\ra}{\right\rangle}
\newcommand{\lb}{\left (}
\newcommand{\rb}{\right )}
\newcommand{\lsb}{\left [}
\newcommand{\rsb}{\right ]}
\newcommand{\beq}{\begin{equation}}
\newcommand{\eeq}{\end{equation}}
\newcommand{\om}{\omega}
\newcommand{\CM}{{\mathcal M}}
\newcommand{\br}{\begin{remark}\rm\small}
\newcommand{\er}{\end{remark}}
\newcommand{\bt}{\begin{theorem}}
\newcommand{\et}{\end{theorem}}
\newcommand{\bd}{\begin{definition}}
\newcommand{\ed}{\end{definition}}
\newcommand{\bl}{\begin{lemma}}
\newcommand{\el}{\end{lemma}}
\newcommand{\bc}{\begin{corollary}}
\newcommand{\ec}{\end{corollary}}

\renewcommand{\d}{{{\partial}}}

\DeclareMathOperator{\Aut}{Aut}

\DeclareMathOperator{\Tr}{Tr}
\DeclareMathOperator{\Res}{Res}

\title[Combinatorics of loop equations]{Combinatorics of loop equations for branched covers of sphere}

\author{P.~Dunin-Barkowski}

\author{N.~Orantin}

\author{A.~Popolitov}

\author{S.~Shadrin}

\address{P.~D.-B.:
Max Planck Institute for Mathematics, Vivatsgasse 7, 53111 Bonn, Germany;
Institute for Theoretical and Experimental Physics, 25 Bolshaya Cheryomushkinskaya Ulitsa, Moscow 117218, Russia; and National Research University Higher School of Economics, Laboratory of Mathematical Physics, 20 Myasnitskaya Ulitsa, Moscow 101000, Russia}
\email{ptdbar@mpim-bonn.mpg.de}

\address{N.~O.:  D\'epartement de math\'ematiques, 
Ecole Polytechnique F\'ed\'erale de Lausanne,
CH-1015 Lausanne,
Switzerland}
\email{nicolas.orantin@epfl.ch}

\address{A.~P.: Korteweg-de~Vries Institute for Mathematics, University of Amsterdam, P.~O.~Box 94248, 1090 GE Amsterdam, The Netherlands and Institute for Theoretical and Experimental Physics, 25 Bolshaya Cheryomushkinskaya Ulitsa, Moscow 117218, Russia}
\email{A.Popolitov@uva.nl}

\address{S.~S.: Korteweg-de~Vries Institute for Mathematics, University of Amsterdam, P.~O.~Box 94248, 1090 GE Amsterdam, The Netherlands}
\email{S.Shadrin@uva.nl}

\begin{document}

\begin{abstract}
We prove, in a purely combinatorial way, the spectral curve topological recursion for the problem of enumeration of bi-colored maps, which
%RED
%in a certain way generalize the notion of
are dual objects to
dessins d'enfant. Furthermore, we give a proof of the quantum spectral curve equation for this problem.
%RED --- Should we remove the following sentence, since the referee doesn't think that this is significant? --- Petya
%As a corollary, we prove a recent conjecture due to Do and Manescu on enumeration of hypermaps and give a new proof for their quantum spectral curve result in this case.
Then we consider the generalized case of 4-colored maps and
%RED
%prove
outline the idea of the proof of
the corresponding spectral curve topological recursion.
\end{abstract}

\maketitle

\tableofcontents

\section{Introduction}

In this paper we discuss the enumeration of \emph{bi-colored maps}. They are decompositions of closed two-dimensional surfaces into polygons of black and white color glued along their sides, considered as combinatorial objects. We count such decomposition of two-dimensional surfaces into a fixed set of polygons with some appropriate weights. This problem is then equivalent to enumeration of Belyi functions with fixed type of local monodromy data over its critical values (following \cite{DoManescu}, we call such functions \emph{hypermaps} in this paper), which is a special case of a more general Hurwitz problem. 

Belyi functions are objects of principle importance in algebraic geometry; they allow to detect the algebraic curves defined over the field of algebraic numbers. There is a way to study them in terms of ``dessins d'enfants'', that is, some embedded graphs in two-dimensional surfaces,  see~\cite{LandoZvonkin} for a survey or~\cite{Dessins} for some recent developments.

The local monodromy data of a Belyi function can be controlled by the choice of three partitions of the degree of the function. We consider a special generating function  for enumeration of Belyi functions. Namely, we fix the length of the first partition to be $n$ and we introduce some formal variables $x_1,\dots,x_n$ to control the first partition as an $n$-point function; we introduce auxiliary parameters $t_i$, $i\geq 1$, in order to control the number of parts of length $i$ in the second partition as a generating function; and we take the sum of all possible choices of the third partition so that the genus of the surface is equal to $g\geq 0$. This way we get some functions $W_{n}^{(g)}(x_1,\dots,x_n)$ that also depend on formal parameters $t_i$, $i\geq 0$. 

As soon as we get some meaningful combinatorial problem, where it is natural to arrange the answers into the generating functions of this type, it makes sense to check whether these functions $W_{n}^{(g)}(x_1,\dots,x_n)$ can be reproduced via the \emph{topological recursion} \cite{EO}. The theory of topological recursion has initially occurred as a way to solve a set of  loop equations satisfied by the correlation functions of a particular class of matrix models~\cite{E1MM, EC1MM,EynardOrantin2MM,ChekhovEynardOrantin}. Then it has evolved to a more abstract and much more general mathematical theory that associates some functions ${\mathcal W}_{n}^{(g)}(x_1,\dots,x_n)$ to some small input related to an algebraic curve called \emph{spectral curve}, see~\cite{EO}. The question is whether we can prove the topological recursion for the generating functions $W_{n}^{(g)}(x_1,\dots,x_n)$ and, if yes, what would be the spectral curve in this case.

For an expert in matrix models the answer is obvious. Indeed, we go back to the original formulation in terms of bi-colored maps. It is a standard representation of correlation functions of a two matrix model, see a survey in~\cite{EynardSurvey} or more recent paper~\cite{AMMN}, and the topological recursion in this case is derived in~\cite{ChekhovEynardOrantin}. However, the general question that one can pose there is whether there is any way to relate the topological recursion to the intrinsic combinatorics of bi-colored maps. There are two steps of derivation of the topological recursion in~\cite{ChekhovEynardOrantin}. First, using skillfully chosen changes of variables in the matrix integral, one can define the \emph{loop equations} for the correlation functions \cite{Eyn2MM}. Then, via a sequence of formal computations, one can determine the spectral curve and prove the topological recursion.

The loop equations of a formal matrix model are equivalent to some combinatorial properties of bi-colored maps~\cite{Tuttesurvey}. 
In this paper, we exhibit these combinatorial relations deriving the loop equations directly from the intrinsic combinatorics of the bi-colored maps. This procedure can be generalized for deriving combinatorially the loop equations of an arbitrary formal matrix model.
 This allows us to give a new, purely combinatorial proof of the topological recursion for the functions $W_{n}^{(g)}(x_1,\dots,x_n)$.
 
%RED
Let us stress that in \cite{EynardSurvey,AMMN,ChekhovEynardOrantin} this problem of counting dessins d'enfant was addressed in matrix model approach. In our paper, by proving loop equations in a combinatorial way,
%and then only using formal manipulations with formal matrix integrals
we have a purely combinatorial approach to this problem.

%RED
Let us also note that although the above mentioned papers dealt with the same numbers (counting dessins d'enfant), different generating functions were considered. The link to the spectral curve topological recursion was not established there. Since spectral curve topological recursion arose in the context of matrix models, the step from a matrix model for a particular counting problem to the topological recursion is a well-known one, and in this particular case it follows from existing works. We stress, however, that in our paper we circumvent the matrix model approach and obtain a purely combinatorial proof for the spectral curve topological recursion.

%RED
%Note that throughout the paper we use the notation from the matrix model theory. We use it only formally, though; all matrix integrals are treated in a formal way and in the end acquire precise combinatorial meaning.

As a motivating example for our work, we use a recent
%RED
%conjecture
question
posed by Do and Manescu in~\cite{DoManescu}. They considered the enumeration problem for  a special case of our bi-colored maps, where all polygons of the white color have the same length $a$. In this case, they conjectured that this enumeration problem satisfies the topological recursion and proposed a particular spectral curve. So, as a special case of our result, we prove their conjecture, and it appears to be a purely combinatorial proof. Though similar problems were considered a lot recently~\cite{KZ,AC1,AC2}, the
%RED
%conjecture of
question posed by
Do and Manescu was not covered there.

There is a general principle that associates to a given spectral curve its quantization, which is a differential operator called \emph{quantum spectral curve}~\cite{GS}. Conjecturally, this operator should annihilate the wave function, which is, roughly speaking, the exponent of the generating series of functions $\int^x\cdots\int^x W_{n}^{(g)}(x_1,\dots,x_n) dx_1\cdots dx_n$. We show that this general principle works in this case, namely, we derive the quantum spectral curve directly from the same combinatorics of loop equations. This generalizes the main result in~\cite{DoManescu} .

The combinatorics that we use in the analysis of bi-colored maps is in fact of a more general nature. The same idea of derivation of the loop equations can be used in more general settings. In particular, we
%RED
%show
outline the idea of
how it would work for the enumeration of 4-colored maps, where the topological recursion was derived from the loop equations by Eynard in~\cite{EynardChain}.

\subsection{Organization of the paper}
In Section \ref{section:branched} we recall the definitions   of hypermaps and discuss generating functions corresponding to hypermap enumeration problems.

%RED
In Section \ref{section:mapsmatrix} we reformulate the definition of hypermaps in terms of bi-colored maps and, for use as a motivation for our combinatorial proof, recall the 2-matrix model which gives rise to enumeration of bi-colored maps.
%RED
%note that in this section all matrix integrals are to be treted formally, they acquire precise combinatorial meaning with the help of the results of the subsequent section.

In Section \ref{section:loopeqs} we recall the form of the loop equations for the 2-matrix model and then we show that using purely combinatorial argument to prove the basic building blocks of loop equations, we can obtain a purely combinatorial proof of the spectral curve for the enumeration of bi-colored maps.

In Section \ref{section:quantqurve} we review the problem of finding the quantum curve for enumeration of hypermaps.

In Section \ref{section:4case} we outline
%RED
the idea of
the proof of the spectral curve topological recursion for the even further generalization of our problem: the case of 4-colored maps, which corresponds to 4-matrix models.

\section{Branched covers of $\mathbb{P}^1$}\label{section:branched}

\subsection{Definitions}

We are interested in the enumeration of covers of $\mathbb{P}^1$ branched over three points. These covers are defined as follows.

\bd \label{def:Mgmn} Consider $m$ positive integers $a_1,\dots,a_m$  and $n$ positive integers $b_1,\dots,b_n$.  
We denote by 
${\CM}_{g,m,n}(a_1,\dots,a_m | b_1,\dots,b_n )$
the weighted count of branched covers of $\mathbb{P}^1$ by a genus $g$ surface with $m+n$ marked points 
$f\colon \left( {\mathcal S} ; q_1,\dots,q_m ; p_1,\dots,p_n \right)\to \mathbb{P}^1$ such that
\begin{itemize}

\item $f$ is unramified over $\mathbb{P}^1 \backslash \{0,1,\infty\}$;

\item the preimage divisor $f^{-1}(\infty)$ is $a_1 q_1 + \dots a_m q_m$;

\item the preimage divisor $f^{-1}(1)$ is $b_1 p_1 + \dots b_n p_n$;

\end{itemize}
Of course, a cover $f$ can exist only if $a_1+\cdots+a_m=b_1+\cdots+b_n$. In this case $d=b_1+\cdots+b_n$ is called the degree of a cover. 

These covers are counted up to isomorphisms preserving the marked points $p_1,\dots,p_n$ pointwise and covering the identity on $\mathbb{P}^1$. The weight of a cover is equal to the inverse order of its automorphism group.

\ed

\begin{example}
In~\cite{DoManescu} the authors consider the case of
$${\CM}_{g,d/a,n}(a,\dots,a | b_1,\dots,b_n ),$$ and relate this enumeration problem to the existence of a quantum curve.
\end{example}

Since such a branched cover can be recovered just from its monodromy around 0, 1 and $\infty$, it is convenient to reformulate this enumeration problem in different terms.

\bd
Let us fix 
$d\geq 1$, $g\geq 0$, $m\geq 1$, and $n\geq 1$. A hypermap of type
$(g,m,n)$
is a triple of permutations $(\sigma_0,\sigma_1,\sigma_\infty) \in S_d^3$ such that
\begin{itemize}

\item $\sigma_0 \sigma_1 \sigma_\infty = Id$;

\item $\sigma_1$ is composed of $n$ cycles;

\item $\sigma_\infty$ is composed of $m$ cycles.

%\item $\sigma_0$ is composed of $v$ cycles where
%\beq
%v = 2- 2g - n - m +d  .
%\eeq
\end{itemize}

A hypermap is called \emph{connected} if the permutations $\sigma_0$, $\sigma_1$, $\sigma_\infty$ generate a transitive subgroup of $S_d$. A hypermap is called \emph{labelled} if the disjoint cycles of $\sigma_1$ are labelled from 1 to $n$.

Two hypermaps $(\sigma_0,\sigma_1,\sigma_\infty)$ and $(\tau_0,\tau_1,\tau_\infty)$ are equivalent if one can conjugate all the $\sigma_i$'s to obtain the $\tau_i$'s. Two labelled hypermaps are equivalent if in addition the conjugation preserves the labelling.

\ed

By Riemann existence theorem, one has

\bl The number 
${\CM}_{g,m,n}(a_1,\dots,a_m | b_1,\dots,b_n)$ is equal to the weighted count of labelled hypermaps of type $(g,m,n)$ where the cycles of $\sigma_\infty$ have lengths $a_1,\dots,a_m$ and the cycles of $\sigma_1$ have length $b_1,\dots,b_n$. Here the weight of a labelled hypermap is the inverse order of its automorphism group.
\el

\subsection{Generating functions}

In order to compute these numbers, it is very useful to collect them in generating functions. For this purpose, we define:

\bd
Let us fix integer $g\geq 0$ and $n\geq 1$ such that $2g-2+n>0$. We also fix one more integer $a\geq 1$ that will be used to restrict the possible length of cycle in $\sigma_\infty$.

The $n$-point correlation function is defined by
\begin{multline}
\Omega_{g,n}^{(a)}(x_1,\dots,x_n):=\\
\sum_{m=0}^\infty \; \sum_{\substack{1\leq a_1,\dots,a_m \leq a \\ 0 \leq b_1,\dots,b_n}} {\CM}_{g,m,n}(a_1,\dots,a_m | b_1,\dots,b_n) \prod_{i=1}^m t_{a_i} \prod_{j=1}^n b_j x_i^{-b_j-1} .
\end{multline}
It is a function of the variables $x_1,\dots,x_n$ that depends on formal parameters $t_1, \dots , t_a$.
\ed

\begin{remark}
Note that the product
\begin{equation*}
{\CM}_{g,m,n}(a_1,\dots,a_m | b_1,\dots,b_n) \; \prod_{j=1}^n b_j
\end{equation*}
counts the same covers as in Definition \ref{def:Mgmn}, but with an additional choice, for each $i$, of one of the possible $b_i$ preimages of a path from $1$ to $0$ starting at point $p_i$. 
\end{remark}

For later convenience in the definition of the quantum curve, we define the symmetric counterpart of the $n$-point correlation function by (for $(g,n) \neq (0,1)$)
\beq \label{eq:Fgndef}
{\mathcal{F}}_{g,n}^{(a)}(x)  := \int^x \dots \int^x \Omega_{g,n}^{(a)}(x_1,\dots,x_n) dx_1 \dots dx_n
\eeq
The special case $(g,n) = (0,1)$, as usual, includes a logarithmic term:
\beq
{\mathcal{F}}_{0,1}^{(a)}(x)  := \log(x)+\int^x \Omega_{0,1}^{(a)}(x_1) dx_1
\eeq
Then we define the wave function by
\beq \label{eq:wave-fun}
Z^{(a)}(x,\hbar) := \exp \left[
\sum_{g = 0}^\infty \sum_{n=1}^\infty {\hbar^{2g+n-2} \over n!} {\mathcal{F}}_{g,n}^{(a)}(x) 
\right].
\eeq

\begin{remark}
Note that in Do and Manescu's paper \cite{DoManescu} a different definition of $F_{g,n}$ was used, differing by $(-1)^n$, which leads to a different definition of $Z^{(a)}$, and, in turn, to a slightly different quantum spectral curve equation. See more on this in Section \ref{section:quantqurve}.
\end{remark}

\section{Maps and matrix models} \label{section:mapsmatrix}

%NOTE:Inserted part
In the present section we discuss the definition of bi-colored maps 
%RED
and review certain matrix model results for the corresponding counting problem.

%RED
These matrix model results serve as a motivation for our combinatorial proof of the spectral curve topological recursion, which is given in the next section.

%RED
Namely, we recall known matrix integral formulas for the generating functions for bi-colored maps, and then we refer to the known proof of the spectral curve topological recursion corresponding to this matrix model. We note that this latter proof only uses the loop equations of the corresponding matrix model as its input. This allows us to give a new, purely combinatorial proof this spectral curve topological recursion, by proving the loop equations independently in a combinatorial way (in Section \ref{section:loopeqs}). This is the main result of the present paper.
%Note that all matrix integrals and matrix model notations in this section are to be treated formally. In the subsequent section, Section \ref{section:loopeqs}, we give an independent, purely combinatorial proof of the loop equations for the matrix model described in the present section. This provides these formal matrix model notations with a precise combinatorial meaning, and gives rise to a combinatorial proof of the spectral curve topological recursion for the counting problem at hand, which is the main result of the present paper.
%RED
%and a formal matrix model argument on the existence of a topological recursion for them. 
%However, it is possible to prove the topological recursion in a purely combinatorial way without even mentioning any formal matrix model integral representation of the problem
%RED ---Should we have the footnote below? If we claim that the combinatorial proof is our main result, this should probably be put in some other way..--- Petya
%\footnote{It is important to note that the combinatorial derivation is equivalent to the matrix model derivation as it is only a translation of the latter in purely combinatorial arguments. However, it might be easier to understand for some readers not used to the matrix model formalism.}
%.
%This is done in the subsequent section, Section~\ref{section:loopeqs}. 
%/Inserted part

\subsection{Covers branched over 3 points and maps}\label{sechypermaps}

There exists a natural graphical representation of hypermaps (which
%RED
are dual objects to
%generalizes the notion of
dessins d'enfant~\cite{LandoZvonkin} and
%RED
generalize
the construction of \cite{DoManescu}).

Let us now describe how to associate a colored map\footnote{In the following, when referring to a map, we refer to a combinatorial object corresponding to a polygonalisation of a surface. These objects appear naturally in the literature in the context of random matrices and were introduced in physics as part of various attempts to quantize gravity in 2 dimensions and to approach string theory from a discrete point of view.} to any labelled hypermap.

Each independent cycle $\rho_i$ in the decomposition of $\sigma_1 = \rho_1 \rho_2 \dots \rho_n$ is represented by a black $|\rho_i|$-gon  whose corners are cyclically ordered and labelled by the numbers composing $\rho_i$. We glue these black polygons by their corners following $\sigma_0$. Namely, for each disjoint cycle $\rho = (\alpha_1, \dots, \alpha_k)$ of $\sigma_0$, one attaches the corners of black faces labelled by $\alpha_1, \dots \alpha_k$ to a $2k$-valent vertex  such that:
\begin{itemize}

\item Turning around the vertex, one encounters alternatively white and black sectors ($k$ of each) separated by the edges adjacent to the vertex;

\item when turning counterclockwise around the vertex starting from the corner labelled by $\alpha_1$, the labels of the corner corresponding to the black sectors adjacent to the vertex form the sequence $\alpha_1, \alpha_2, \dots, \alpha_k$.

\end{itemize}

\begin{example}
Let us give an example of a bi-colored map. Consider a hypermap  corresponding to $d=7$, $g=0$,
\begin{align*}
\sigma_{0} &= (1,5,7)(4,6), \\
\sigma_{1} &= (1,2,3,4)(5,6,7), \\
\sigma_{\infty} &= (1,6,3,2)(4,5) 
\end{align*}
Then the corresponding bi-colored map can be seen in Figure \ref{colmapfig}.
\begin{figure}[H]
\begin{tikzpicture}
[
dott/.style={circle,draw=black,fill=black,inner sep=0pt,minimum size=5pt}
]
\draw [fill=black!15] (2,-1.5) .. 
 controls (-2,-1.5) and (2,-1.5) .. (-2,-1.5) ..
 controls (-2,2.5) and (-2,-1.5) .. (-2,2.5) .. 
 controls (2,2.5) and (-2,2.5) .. (2,2.5) .. 
 controls (1,4.5) and (-2,4.5) .. (-3,4) ..
 controls (-4,3.5) and (-4.5,2) .. (-4.5,0.5) ..
 controls (-4.5,-1) and (-4,-2.5) .. (-3,-3) .. 
 controls (-2,-3.5) and (1,-3.5) .. (2,-1.5);

\draw [fill=black!15] (2,2.5) .. 
controls (2,-1.5) and (2,2.5) .. (2,-1.5) .. 
controls (5,-1.5) and (5.5,-2) .. (5,-2.5) .. 
controls (4.5,-3) and (2.5,-2) .. (2,-1.5) .. 
controls (2,-4) and (3,-4) .. (4,-4) .. 
controls (5.5,-4) and (6.5,-3.5) .. (6.5,-2) .. 
controls (6.5,1.5) and (5,2.5) .. (2,2.5);

\node [dott,label={[label distance=5pt]170:4},
  label={[label distance=3pt]300:6}] at (2,2.5) {};

\node [dott,label={[label distance=5pt]170:3}] at (-2,2.5) {};

\node [dott,label={[label distance=5pt]-170:2}] at (-2,-1.5) {};

\node [dott,label={[label distance=5pt]-170:1},
  label={[label distance=5pt]275:5},
  label={[label distance=3pt]45:7}] at (2,-1.5) {};
\end{tikzpicture}
\caption{Bi-colored map}
\label{colmapfig}
\end{figure}
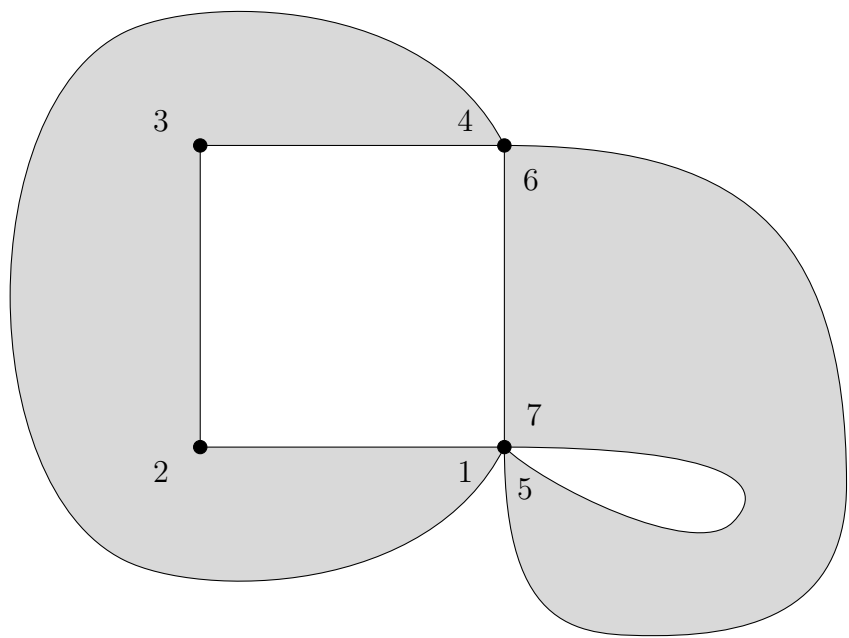
In this figure we see two black polygons corresponding to cycles $(1,2,3,4)$ and $(5,6,7)$ of $\sigma_1$; they are glued according to $\sigma_0$.
\end{example}

Let us fix $a\geq 1$. We denote by $G_{g,m,n}^{(a)}$ the set of bi-colored maps, where $m$ is the number of white polygons, $n$ is the number of black polygons, and $g$ is the genus of the surface we get by gluing the polygons and $a$ is the maximum perimeter of a white polygon. We assume that the black polygons are labelled, and we consider the maps up to combinatorial isomorphisms preserving this labelling. For a particular map $M\in G_{g,m,n}^{(a)}$ we denote by $\Aut(M)$ its automorphism group.

One can restate the problem of enumerating covers of $\mathbb{P}^1$ as counting bi-colored maps as follows.

\begin{lemma} The function $\Omega_{g,n}^{(a)}(x_1,\dots,x_n)$ is the generating function of bi-colored maps with an arbitrary number $m\geq 1$ of white faces whose perimeters are less or equal to $a$ and $n$ marked black faces with perimeters $b_1,\dots,b_n$. That is,
\beq \label{eq:bicoloredgen}
\Omega_{g,n}^{(a)}(x_1,\dots,x_n) = \sum_{m=1}^\infty \sum_{M \in G_{g,m,n}^{(a)}} \frac{\prod_{i=1}^a t_i^{n_i(M)}}{\left|\Aut(M)\right|} \prod_{j=1}^n b_j(M) x_j^{-b_j(M)-1} .
\eeq
Here by $n_i(M)$ we denote the number of white polygons of perimeter $i$ in $M$, and $b_1(M),\dots,b_n(M)$ are the perimeters of the black polygons in $M$. 
\end{lemma}

\subsection{Matrix model and topological recursion}

%RED
In the present subsection we recall the matrix model techniques of solving the problem of enumeration of bi-colored maps, which provide the motivation for our subsequent combinatorial proof of spectral curve topological recursion for this problem.

The enumeration of bi-colored maps is a classical problem of random matrix theory which is equivalent to the computation of formal matrix integrals. One can state this equivalence in the following way.
\bl
(see, e.~g. \cite{EynardSurvey})
Consider the partition function of a formal Hermitian two-matrix model 
\begin{align}
& {\mathcal{Z}}\left(\vec{t}^{(1)}, \vec{t}^{(2)} \right) := 
\\ \notag
& \int_{H_N}^{formal} dM_1 \, dM_2 \, e^{-{N
%\over t
} \left[ \Tr(M_1 M_2) -  \Tr V_1(M_1) - \Tr V_2(M_2) \right]}
\end{align}
where the potentials $V_i(x)$, $i=1,2$, are polynomials of degree $d_i$,
\beq
V_i(x) = \sum_{d=1}^{d_i} \frac{t_d^{(i)} }{ d} x^d.
\eeq
This partition function is a generating function of bi-colored maps, that is,
\begin{align}
&
{\mathcal{Z}}\left(\vec{t}^{(1)}, \vec{t}^{(2)} \right) = 
\\ \notag & 
\sum_{g,m,n=0}^\infty \sum_{M \in {\mathcal S}^\bullet_{g,m,n}}\frac { \prod\limits_{i=1}^{d_1} \left[t_i^{(1)} \right]^{n_i^{(1)}(M)} \prod\limits_{i=1}^{d_2} \left[t_i^{(2)} \right]^{n_i^{(2)}(M)}
%\prod\limits_{k,l=1}^2 \left[C^{-1}\right]_{i,j}^{n_{i,j}(M)}
}{\left|\Aut(M) \right|}
\end{align}
where
\begin{itemize}
\item ${\mathcal S}^\bullet_{g,m,n}$ is the set of bi-colored maps, possibly disconnected, of genus $g$ composed of $n$ black polygons and $m$ white polygons glued by their edges, such that black polygons are glued only to white polygons and vice versa. Neither black nor white polygons are marked. 
\item By $n_i^{(1)}(M)$ (resp.~$n_i^{(2)}(M)$) we denote the number of black (resp. white) polygons of perimeter $i$ in $M$;
\end{itemize}
\el

It is also possible to enumerate connected maps with some specific marked faces by computing certain correlation functions of this formal matrix model.

\bd
For any set of words (non-commutative monomials) $\{f_i(x,y)\}_{i=1}^s$ in two variables,
we define the correlator of the formal matrix model by
\begin{equation*}
\left<{\displaystyle \prod_{i=1}^s} \Tr f_i (M_1,M_2)\right>:= 
\frac{ \int_{H_N}^{formal} d\mu_N(M_1,M_2)\, \prod_{i=1}^s \Tr f_i (M_1,M_2)}{{\mathcal{Z}}\left(\vec{t}^{(1)}, \vec{t}^{(2)}
%, C
 \right)},
\end{equation*}
where the measure of integration $\mu(M_1,M_2)$ is the same as before, 
\begin{equation*}
d\mu_N(M_1,M_2):=dM_1\, dM_2\,  e^{-{N
%\over t
} \left[
%\sum\limits_{i,j=1}^2 C_{ij} \Tr(M_i M_j)
\Tr(M_1 M_2)
-  \Tr V_1(M_1) - \Tr V_2(M_2) \right] }.
\end{equation*}
We denote by $\left<{\displaystyle \prod_{i=1}^s} \Tr f_i (M_1,M_2)\right>_c$ its connected part.

\ed

In matrix models, one classically works with generating series of such correlators (named correlation functions) defined by
$$
W_{k,l}(x_1,\dots,x_k;y_1,\dots,y_l):=
%N^{k+l-2}
\left<\prod_{i=1}^k \Tr {1 \over x_i-M_1} \prod_{j=1}^l \Tr{ 1 \over y_j -M_2} \right>_c .
$$
These correlation functions have to be understood as series expansions around $x_i,y_i \to \infty$:
\begin{equation} \label{eq:Wklseries}
W_{k,l}(x_1,\dots,x_k;y_1,\dots,y_l):=
%N^{k+l-2}
\sum_{\vec n \in \mathbb{N}^k} \sum_{\vec m \in \mathbb{N}^l} \left<\prod_{i=1}^k {\Tr M_1^{n_i} \over x_i^{n_i+1}} \prod_{j=1}^l {\Tr M_2^{m_j} \over y_j^{m_j+1}} \right>_c .
\end{equation}
These correlation functions admit a topological expansion, i.~e. they can be written as 
$$
W_{k,l}(x_1,\dots,x_k;y_1,\dots,y_l) = \sum_{g=0}^\infty N^{2-2g-k-l} W_{k,l}^{(g)}(x_1,\dots,x_k;y_1,\dots,y_l)
$$
where each of $W^{(g)}_{k,l}$ does not depend on $N$.

With this notation,
\begin{align*}
&
W^{(g)}_{k,l}(x_1,\dots,x_k;y_1,\dots,y_l) = 
\\ & \phantom{privet}
\sum_{m,n=0}^{\infty}
\sum_{\substack{\vec\alpha \in \mathbb{N}^k \\ \vec\beta \in \mathbb{N}^l}}
\sum_{M \in {\mathcal S}^\circ_{g,m,n|\vec \alpha, \vec \beta}} 
 \frac
  {\prod\limits_{i=1}^{d_1} \left[t_i^{(1)} \right]^{n_i^{(1)}(M)} \prod\limits_{i=1}^{d_2} \left[t_i^{(2)} \right]^{n_i^{(2)}(M)} %\prod\limits_{k,l=1}^2 \left[C^{-1}\right]_{i,j}^{n_{i,j}(M)} 
	}{
	\left|\Aut(M) \right| \,  \prod\limits_{i=1}^k x_i^{\alpha_i+1}  \prod\limits_{j=1}^l y_j^{\beta_j+1}
},
\end{align*}
where ${\mathcal S}^\circ_{g,m,n|\vec \alpha, \vec \beta}$ is the set of connected bi-colored maps of genus $g$ composed of $n$ unmarked black faces, $m$ unmarked white faces, $k$ marked black faces of perimeters $\alpha_1,\dots,\alpha_k$, each having one marked edge, and $l$ marked white faces of perimeter $\beta_1,\dots,\beta_n$, each having one marked edge too; black faces are only glued to white faces and vice versa, as above.
%In particular, this means that
%$$
%\Omega_{g,n}^{(a)}(x_1,\dots,x_n)
% =\left.  W^{(g)}_{n,0}(x_1,\dots,x_n;\emptyset)\right|_{t_i^{(1)} = 0\ \mathrm{for}\  i=1,\dots,d_1} 
%$$
%if we assume that  $d_2=a$ and $C_{1,1}=C_{2,2}=0$, $C_{12}=C_{21}=1$.

Such a model admits a spectral curve. This means that there exists a polynomial $P(x,y)$ of degree $d_1-1$ in $x$ and $d_2-1$ in $y$ such that the generating function for discs $W_{1,0}^{(0)}(x)$ satisfies an algebraic equation:
$$
E_{2MM}(x,W_{1,0}^{(0)}(x)) = 0, \qquad x\in\mathbb{C},
$$
where
$$
E_{2MM}(x,y) = (V_1'(x) - y) (V_2'(y) - x) -P(x,y) +1 .
$$
In \cite{EynardOrantin2MM,ChekhovEynardOrantin}, it was proved that the correlation functions $W_{k,0}^{(g)}$ can be computed by topological recursion on this spectral curve.

%RED
First, let us recall the definition of the spectral curve topological recursion.
\bd \cite{EO}
Let $C$ be an algebraic curve, $x$ and $y$ two functions on $C$ and $\om_{0,2}$ a bidifferential defined on $C\times C$. Denote $ydx$ by $\om_{0,1}$ and let $a_i$ stand for zeroes of $dx$ and $\sigma_i(z)$ stand for the deck transformation near $a_i$.
Then \emph{spectral curve topological recursion} defines $n$-multidifferentials $\om_{g,n}$ by the following recursive formula 
\begin{multline}
\label{eq:CEO}
\om_{g,n}(z_1,\dots,z_n)\\
=
\frac{1}{2} \sum_{i} \mathop{\Res}_{z->a_i} 
\frac{\int_z ^{\sigma_i(z)} \om_{0,2}(\;\cdot\;,z_1)}
{\om_{0,1}(\sigma_i(z)) - \om_{0,1}(z)}
\Bigg[\om_{g-1,n+1}(z,\sigma_i(z),z_2,\dots,z_n)
\\
+
\sum_{\substack{g_1+g_1=g\\
I\sqcup J=\{2,\dots,n\}}} ^{\text{stable}}
\om_{g_1,|I|+1}(z,z_I)\om_{g_2,|J|+1}
(\sigma_i(z),z_J)
\Bigg],
\end{multline}
"Stable" above the summation sign stands for taking the sum excluding the terms where ($g_1,|I|) = (0,1)$ or ($g_2,|J|) = (0,1)$.
\ed

\bt \cite{EynardOrantin2MM,ChekhovEynardOrantin}
The correlation functions of the 2-matrix models can be computed by the topological recursion procedure of \cite{EO} with the genus $0$ spectral curve
$$
E_{2MM}(x,y) = (V_1'(x) - y) (V_2'(y) - x) -P(x,y) +1 
$$
and the genus $0$, 2-point function defined by the bilinear differential
$$
\om_{0,2}(z_1,z_2) = {dz_1 dz_2 \over (z_1-z_2)^2} 
$$
for a global coordinate $z$ on the spectral curve.
\et
The proof of this theorem consists in three steps:
\begin{itemize}
\item First, find a set of equations satisfied by the correlation functions of the matrix model.

\item Second, show that these equations admit a unique solution admitting a topological expansion.

\item Third, exhibit a solution which immediately implies the topological recursion.

\end{itemize}

\subsection{A matrix model for branched covers}

Since the problem of enumerating branched covers can be rephrased in terms of bi-colored maps, one can find a matrix model representation for it.

Using the definition of the preceding section together with the hypermap representation of section \ref{sechypermaps}, one immediately finds that
\bl
The correlation functions of the formal two matrix model with
 potentials $V_1(x) = 0$ and $V_2(x) = {\displaystyle \sum_{i=1}^a} {t_i \over i} x^{i} $
 %, and interaction matrix
 %\beq
 %C = \dfrac{1}{2}\,\left[ \begin{array}{cc}
 %0 & 1 \cr
 %1 & 0 \cr
 %\end{array}
 %\right]
 %\eeq
coincide with the generating series of covers of $\mathbb{P}^1$ branched over 3 points defined in \eqref{eq:bicoloredgen}, for $(g,k)\neq(0,1)$:
\beq \label{eq:WOmegaident}
W_{k,0}^{(g)}(x_1,\dots,x_k) = \Omega_{g,k}^{(a)}(x_1,\dots,x_k).
\eeq
For $(g,k)=(0,1)$ we have
\beq 
W_{1,0}^{(0)}(x_1) = \dfrac{1}{x}+\Omega_{0,1}^{(a)}(x_1).
\eeq
\el

Applying \cite{EO,ChekhovEynardOrantin}, one can thus compute the generating series using topological recursion.

We have:
%RED
%\bc
\bt\label{corotoporec}
The generating series for hypermaps
%RED
%$ \Omega_{g,k}^{(a)}(x_1,\dots,x_k)$
\begin{equation}
\om_{g_n}(x_1,\dots,x_k) =  \Omega_{g,k}^{(a)}(x_1,\dots,x_k)dx_1\dots dx_n 
\end{equation}
can be computed by topological recursion with a genus 0 spectral curve 
\beq \label{eq:classspectcurve}
E^{(a)}(x,y) = y\,  \left( \sum_{i=1}^a t_i y^{i-1} -x\right) +1 = 0
\eeq
and the genus 0 2-point function defined by the corresponding Bergmann kernel, i.~e.
\beq
\om_{0,2}(z_1,z_2) = {dz_1 \otimes dz_2 \over (z_1-z_2)^2}
\eeq
for a global coordinate $z$ on the genus 0 spectral curve.

%RED
%\ec
\et

%RED
For brevity, we are not reproducing here the arguments from \cite{EO,ChekhovEynardOrantin}, but we note that the only fact about the matrix model that is used in these arguments to prove the spectral curve topological recursion is the loop equations for the matrix model. 

In the next section we prove these loop equations independently in a combinatorial way, and thus obtain a new, purely combinatorial, proof of Theorem \ref{corotoporec}, which is the main result of the present paper.

\br
Theorem \ref{corotoporec}
%RED
%proves in particular the conjecture made
in particular answers the question
by Do and Manescu \cite{DoManescu} considering such covers with only type $a$ ramifications above 1. The spectral curve is indeed, like Do and Manescu suggested,
\beq
E^{(a)}(x,y) = y\,  \left(y^{a-1} -x\right) +1 = 0
\eeq
coinciding with the classical limit of their quantum curve.

\er

%\br
%We can also use a small modification of the formal two matrix model, with the measure 
%\begin{equation*}
%d\mu_N(M_1,M_2):=dM_1\, dM_2\,  e^{-{N
%		\over t
%	} \left[
%	%\sum\limits_{i,j=1}^2 C_{ij} \Tr(M_i M_j)
%	\Tr(M_1 M_2)- \Tr V_2(M_2) \right] },
%\end{equation*}
%where $V_2(x) = { \sum_{i=1}^\infty} {x^{i} \over i}  $. In this case, though $V_2$ is no longer polynomial, we can use expansion in $t$ in order to restrict the degree 
%
%\er

\section{Loop equations and combinatorics} \label{section:loopeqs}

The proof of Theorem~\ref{corotoporec} in \cite{EO,ChekhovEynardOrantin} relies on the representation of our combinatorial objects in the form of a formal matrix integral.
Actually, the only input from the formal matrix model is the existence of loop equations satisfied by the correlation functions of the model. These loop equations are of combinatorial nature and should reflect some cut-and-join procedure satisfied by the hypermaps being enumerated. 
However, a simple combinatorial interpretation of these precise 2-matrix model loop equations could not be found in the literature, even if some similar and probably equivalent equations  have been derived combinatorially in some particular cases \cite{BBM,Tuttesurvey}. In this section, we derive such an interpretation, allowing to bypass the necessity to use any integral (matrix model) representation and thus getting a completely combinatorial proof of the results of the preceding section.

\br
While writing the paper, we have been informed that such a direct derivation of the loop equations for the 2-matrix model is performed in chapter 8 \cite{Eynbook} which is in preparation and whose preliminary version can be found online.

\er

\subsection{Loop equations}

In order to produce the hierarchy of loop equations whose solution gives rise to the topological recursion, one combines two set of equations which can be written as follows:

\begin{itemize}

\item The first one corresponds to the change of variable
\begin{equation*}
M_2 \to M_2 + \epsilon { 1 \over x-M_1} {\displaystyle \prod_{i=1}^n} \Tr {1 \over x_i-M_1}
\end{equation*}
in the formal matrix integral defining the partition function. To first order in $\epsilon$, the compensation of the Jacobian (which is vanishing here) with the variation of the action gives rise to the equation:
\begin{align} \label{eq:loop1}
& %(C_{1,2} + C_{2,1})
\left<\Tr \left({ M_1 \over x-M_1}\right) {\displaystyle  \prod_{i=1}^n} \Tr {1 \over x_i-M_1} \right>  
\\ \notag & 
%+ 2 C_{2,2}  \left<\Tr \left({ 1 \over x-M_1} M_2 \right) {\displaystyle  \prod_{i=1}^n} \Tr {1 \over x_i-M_1} \right> 
\\ \notag &
=  \left<\Tr \left({ 1 \over x-M_1} V_2'(M_2) \right) {\displaystyle  \prod_{i=1}^n} \Tr {1 \over x_i-M_1} \right>
\end{align}

\item The second one corresponds to the change of variable
\beq
M_1 \to M_1 + \epsilon {1 \over x-M_1}{V_2'(y) - V_2'(M_2) \over y-M_2} \prod_{i=1}^n \Tr {1 \over x_i-M_1}
\eeq
and reads
\begin{align} \label{eq:loop2}
& 
%{2 C_{1,1} } \left< \Tr \left({M_1 \over x-M_1}{V_2'(y) - V_2'(M_2) \over y-M_2}\right) {\displaystyle \prod_{i=1}^n} \Tr {1 \over x_i-M_1}  \right> 
%\\ \notag &
%+ (C_{1,2} + C_{2,1})
\left< \Tr \left({1 \over x-M_1}{V_2'(y) - V_2'(M_2) \over y-M_2} M_2 \right) {\displaystyle \prod_{i=1}^n} \Tr {1 \over x_i-M_1}  \right>   
\\ \notag &
 = \left< \Tr \left({V_1'(M_1) \over x-M_1}{V_2'(y) - V_2'(M_2) \over y-M_2}  \right) {\displaystyle \prod_{i=1}^n} \Tr {1 \over x_i-M_1}  \right> 
\\ \notag &
 + {1 \over N} \left< \Tr \left( {1 \over x-M_1} \right) \Tr \left({1 \over x-M_1}{V_2'(y) - V_2'(M_2) \over y-M_2} \right) {\displaystyle \prod_{i=1}^n} \Tr {1 \over x_i-M_1}  \right>
\\ \notag &
 + \frac {1 }{ N} {\displaystyle \sum_{i=1}^n} \left<\Tr \left( {1 \over (x_i-M_1)^2} {1 \over x-M_1} {V_2'(y) - V_2'(M_2) \over y-M_2} \right) {\displaystyle \prod_{j\neq i}} \Tr {1 \over x_j-M_1}  \right>
\end{align}

\end{itemize}

Note that in these equations the correlators are not the connected ones, but they are generating functions of possibly disconnected maps of arbitrary genus.

\subsection{Combinatorial interpretation}

The loop equations~\eqref{eq:loop1}, \eqref{eq:loop2} make sense only in their $x,x_i,y \to \infty$ series expansions. These expansions generate a set of equations for the correlators of the matrix models which can be interpreted as relations between the number of bi-colored maps with different boundary conditions. In this section, we give a combinatorial derivation of these relations.

\subsubsection{Definition of boundary conditions}

In order to derive the loop equations, we have to deal with bi-colored maps with boundaries (or marked faces) of general type. A map with $n$ boundaries is a map with $n$ marked faces (polygons), each carrying a marked edge. The boundary conditions are defined as the color of the marked face.

However, in the following, we need to also introduce mixed-type boundary conditions described as follows.

%First, let us introduce the notion of a \emph{side} of an edge of the map. 
%obtained by considering marked faces with sides colored differently.
%RED
A bi-colored map with $n$ mixed-type boundaries is a map with $n$ marked faces (each with a marked edge), where all unmarked faces are colored either black or white (as usual, black faces can border only white ones and vice versa). The marked faces are uncolored. In addition, edges of the map are colored in the following way. We say that each edge has two \emph{flanks}, associated with two possible normal directions to the edge. Each of these two flanks for each edge is colored either black or white such that 1) for a given edge its two flanks are oppositely colored and 2) if a given edge belongs to an unmarked face, its flank in the direction of this face has the same color as the face. For convenience, for a given face $\mathcal{F}$ let us call the $\mathcal{F}$-facing flanks of the edges of $\mathcal{F}$ \emph{inner} with respect to $\mathcal{F}$, and the opposite flanks \emph{outer}.

%We however, assign color to the \emph{edges} of marked faces (color of an edge of a marked face is a property of the given edge as part of the given face, ), such that black edge borders a white unmarked face and vice versa.

%RED
The boundary conditions of a marked face are then given by the sequence of colors of the inner flanks of the edges of this face starting from the marked edge and going clockwise from it.

%RED
For a given marked face consider a set of $n$ sequences of non-negative integers
\beq
S_i = b_{i,1}, a_{i,1}, b_{i,2} , a_{i,2} \dots b_{i,l_i}, a_{i,l_i} , \qquad i=1,\dots n. 
\eeq
Here $b_{i,1}$ is the number of consecutive inner black flanks starting from the inner flank of the marked edge and going clockwise (it is equal to zero if the marked edge is white), $a_{i,1}$ is the number of the following consecutive white flanks, and so on. 

We define ${\mathcal T}_{S_1,\dots,S_n}^{(g)}$ to be the number of connected bi-colored maps of genus $g$ with $n$ boundaries with the boundary conditions $S_1,\dots, S_n$.

\br
In terms of correlators of a two matrix model, one can write
\beq
{\mathcal T}_{S_1,\dots,S_n}^{(g)} = N^{n+2g-2}
\left<
\prod_{i=1}^n \Tr
%\left(M_1^{a_{i,1}} M_2^{b_{i,1}} M_1^{a_{i,2}} M_2^{b_{i,2}} \dots M_1^{a_{i,n}} M_2^{b_{i,n}} \right)
\left(M_1^{b_{i,1}} M_2^{a_{i,1}} M_1^{b_{i,2}} M_2^{a_{i,2}} \dots M_1^{b_{i,l_i}} M_2^{a_{i,l_i}} \right)
\right>_c^g
\eeq
where the superscript $g$ means that we only consider the $g$'th term of the expansion in $N^{-2}$ of this correlator.

\er

\subsubsection{Cut-and-join equations}

With these definitions, we are ready to derive the loop equations \eqref{eq:loop1} and \eqref{eq:loop2}.

%First, let us note that, due to the presence of diagonal terms in matrix $C^{-1}$, on the other side of some given edge $e$ of a marked face we can have a polygon not only of the opposite color to the color of $e$ but also of the same color. For all such cases we recolor $e$ into the opposite color and add weight 
%\begin{equation}
%\dfrac{[C^{-1}]_{11}}{[C^{-1}]_{12}}
%\end{equation}

%A remark is in order concerning the interpretation of the interaction terms coming from the matrix $\left[C^{-1}\right]$. Its diagonal terms weight the gluing of two polygons of the same color along one of their edges. We can always interpret these terms as mono-colored 2-gons and consider them as a shift of the quadratic terms of the potential. With this interpretation, one enumerates only bipartite maps where only polygons of different colors can be glued together. The weight of the 2-gons of color $i$ is then shifted by $C_{ii}$. This is the interpretation that we consider below.

%\br
%By doing so, one inverts the weight of the gluing, the 2-gons being weighted by $C$ and not its inverse.
%However it is important to notice that one did not change the enumeration problem since such a 2-gon can be glued only using propagators changing colors, it always comes between two of these, getting a weight $[C^{-1}]_{ij} C_{jj}[C^{-1}]_{ji}$ leading to the expected weight $[C^{-1}]_{ii}$.

%One should also notice that two 2-gons cannot be glued together in this representation.
%\er

Namely, we can generalize to the two matrix model the procedure developed by Tutte for the enumeration of maps \cite{Tutte} and then extensively developed in the study of formal random matrices. Let us consider a connected genus $g$ map with $n+1$ boundaries with boundary conditions
\beq
S_0 = k+1,0; \qquad S_i = k_i,0, \quad i=1,\dots,n.
\eeq
%RED
This means that the inner flanks of all the edges of the marked faces are black.
This map contributes to ${\mathcal{T}}_{k+1,0;k_1,0;\dots;k_n,0}^{(g)}$.
Let us remove the marked edge from the boundary $0$. Since one can only glue together faces of different colors, on the other side of the marked edge one can find only a white (unmarked) $l$-gon with $1\leq l \leq d_2$. After removing the edge, let us mark in the resulting joint polygon the edge which is located clockwise from the origin of the removed edge (the origin of an edge is the vertex located on the counterclockwise side of the edge). We end up with a map that contributes to ${\mathcal{T}}_{0,l-1,k,0;k_1,0;\dots;k_n,0}^{(g)}$. This procedure is bijective between the sets considered. We take the sum over all possibilities, taking into account the weight of the edge and $l$-gon removed, and we see that
\beq
{\mathcal{T}}_{k+1,0;k_1,0;\dots;k_n,0}^{(g)}  = 
%\left[C^{-1}\right]_{1,2}
\sum_{l=1}^{d_2} t_l^{(2)} {\mathcal{T}}_{0,l-1,k,0;k_1,0;\dots;k_n,0}^{(g)}.
\eeq 
Multiplying by $x^{-k-1}x^{-k_1-1}\dots x^{-k_n-1}$ and taking the sum over\\ $k,k_1,\dots,k_n$, one recovers the loop equation \eqref{eq:loop1}.

This first equation produces mixed boundary condition out of homogeneous
%white
black
conditions. Let us now proceed one step further and apply Tutte's method to the maps produced in this way.

Let us consider a map contributing to ${\mathcal{T}}_{0,l+1,k,0;k_1,0;\dots;k_n,0}^{(g)}$, i.~e. a genus $g$ connected map with boundary condition:
\beq
S_0 = 0,l+1,k,0; \qquad S_i = k_i,0, \quad i=1,\dots,n.
\eeq
Note that it follows from our definition that the inner flank of the marked edge of the $0-th$ marked face is of white boundary condition type. When we remove it, we can produce different types of maps, namely, strictly one of the three following cases takes place. 
\begin{itemize}

%RED
\item On the other side of the edge lies an unmarked black $m$-gon. We remove the edge and this gives a map that contributes to ${\mathcal{T}}_{0,l,k+m-1,0;k_1,0;\dots;k_n,0}^{(g)}$.

%RED
\item The opposite flank of the edge is a black inner flank of the same marked face. Then two possible cases occur. The resulting surface can still be connected, giving rise to a map contributing to ${\mathcal{T}}_{m,0;0,l,k-m,0;k_1,0;\dots;k_n,0}^{(g-1)}$ for some $1\leq m \leq k$, i.~e. with one more boundary but a genus decreased by one. Or removing the marked edge can disconnect the map into two connected component giving contributions to ${\mathcal{T}}_{m,0;k_{\alpha_1},0;\dots;k_{\alpha_j},0}^{(h)}$ and ${\mathcal{T}}_{0,l,k-m,0;k_{\beta_1},0;\dots;k_{\beta_{n-j}},0}^{(g-h)}$ respectively, where $0\leq h \leq g$ and $\{\alpha_1,\dots,\alpha_j\} \cup \{\beta_1,\dots,\beta_{n-j}\} = \{1,\dots,n\}$. This type of behavior can be thought of as a "cut" move.

%RED
\item On the other side of the edge lies a marked black face with boundary condition $(k_i,0)$. Removing the edge, one gets a contribution to ${\mathcal{T}}_{0,l,k+k_i-1,0;k_1,0;\dots;k_{i-1},0;k_i-m,0;k_{i+1},0 \dots;k_n,0}^{(g)}$.  This type of behavior can be thought of as a "join" move.

\end{itemize}
Once again, this procedure is bijective, if we take the sum over all cases. Taking into account the weight of the elements removed, we end up with an equation relating the number of bi-colored maps with different boundary conditions:
\begin{align}
&
%C_{1,2}
{\mathcal{T}}_{0,l+1,k,0;k_1,0;\dots;k_n,0}^{(g)}\\ \notag
&= {\displaystyle \sum_{m=0}^{d_2}} t_m^{(2)} {\mathcal{T}}_{0,l,k+m-1,0;k_1,0;\dots;k_n,0}^{(g)} 
\\ \notag
& + {\displaystyle \sum_{m=0}^k }{\mathcal{T}}_{m,0;0,l,k-m,0;k_1,0;\dots;k_n,0}^{(g-1)} \\ \notag
&+ {\displaystyle  \sum_{m=0}^k \sum_{h=0}^g \sum_{\vec \alpha \cup \vec \beta = \{1,\dots,n\}}} {\mathcal{T}}_{m,0;k_{\alpha_1},0;\dots;k_{\alpha_j},0}^{(h)} {\mathcal{T}}_{0,l,k-m,0;k_{\beta_1},0;\dots;k_{\beta_{n-j}},0}^{(g-h)} \\ \notag
&+ {\displaystyle \sum_{i=1}^n} {\mathcal{T}}_{0,l,k+k_i-1,0;k_1,0;\dots;k_{i-1},0;k_i-m,0;k_{i+1},0 \dots;k_n,0}^{(g)} 
\end{align}
where $\vec \alpha = \{\alpha_1,\dots,\alpha_j\}$ and $\vec \beta = \{\beta_1,\dots,\beta_{n-j}\}$.
This equation is the genus $g$ contribution to the expansion of the loop equation~\eqref{eq:loop2} when all its variables are large.

This concludes the fully combinatorial proof of the two matrix model's loop equations. The latter can be seen as some particular cut-and-join equations. One can now apply the procedure used in \cite{ChekhovEynardOrantin} for solving them (without having to introduce any matrix model consideration!) and derive the topological recursion for the generating functions of bi-colored maps with homogenous boundary conditions,
%RED
which implies Theorem \ref{corotoporec}.

%%%%%%%%
%%%%%%%%
%%%%%%%%

\section{Quantum curve} \label{section:quantqurve}

In this section we prove a generalization of the theorem of  Do and Manescu from~\cite{DoManescu} on the quantum spectral curve equation for enumeration of hypermaps.

\bt \label{thm:quantum-curve}
The wave function $Z^{(a)}(x)$,
defined in
\eqref{eq:wave-fun}, satisfies the ODE:
\begin{align}
\lb -\hbar x \frac{\p}{\p x} + 1 + \sum_{i=1}^a t_i \lb \hbar \frac{\p}{\p x}\rb^i \rb Z^{(a)}(x) = 0
\end{align}
\et
\begin{remark}
The differential operator in the previous theorem is given by the naive quantization of the classical spectral curve \eqref{eq:classspectcurve}, $y \leftrightarrow \hbar \dfrac{\p}{\p x}$. Note that in Do and Manescu's paper \cite{DoManescu} a different definition of $Z^{(a)}$ was used, as noted above, and a different convention $y \leftrightarrow - \hbar \dfrac{\p}{\p x}$.
\end{remark}

\subsection{Wave functions}

In the proof we use the notations coming from the formal matrix model formalism for simplicity, but as usual in the formal matrix model setup, they just represent well defined combinatorial objects which satisfy the loop equations derived in the preceding sections.

In what follows we identify $N$ with $1/\hbar$.

From the definition of the wave function $Z^{(a)}$, given in formulas~\eqref{eq:Fgndef}-\eqref{eq:wave-fun}, from the identification between $W^{(g)}_{k,0}$ and $\Omega^{(a)}_{g,k}$ given by Equation~\eqref{eq:WOmegaident} and from the definition of $W_{k,l}$ \eqref{eq:Wklseries}, we have
\begin{equation}
Z^{(a)}(x) = \exp\left(\dfrac{1}{\hbar}\log(x)+
\sum_{n = 1}^\infty \dfrac{(-1)^n}{n!}
\sum_{b_1,\dots b_n = 1}^\infty
 \frac{\la \Tr(M_1^{b_1})\dots \Tr(M_1^{b_n})\ra_c }{b_1 \dots b_n x^{b_1} \dots x^{b_n}}\right).
\end{equation}
The standard relation between connected and disconnected correlators imply
\begin{equation}
Z^{(a)}(x) = x^{1/\hbar}
\sum_{n = 1}^\infty  \dfrac{(-1)^n}{n!}
\sum_{b_1,\dots b_n = 1}^\infty
 \frac{\la \Tr(M_1^{b_1})\dots \Tr(M_1^{b_n})\ra }{b_1 \dots b_n x^{b_1} \dots x^{b_n}} .
\end{equation}
%where in the correlators we identify $N$ to be equal $1/\hbar$.

In order to simplify the notation, we introduce functions $Z_n^r(x_1,\dots,x_n)$ and $Z^r(y,x)$ for integers $n\geq 1$, $r\geq 0$ (we call these functions non-principally-specialized wave functions).
\definition{The $n$-point wave function $Z^r_n$ of level $r$ is defined as
\begin{align}
&Z^r_n (x_1, \dots , x_n) :=\\ \notag
&\log(x_1)\, (-1)^{n-1} \sum_{b_2 \dots b_n = 1}^\infty
\frac{\la \Tr(M_2^r) \Tr(M_1^{b_2}) \dots \Tr(M_1^{b_n}) \ra}{b_2 \dots b_n\
x_2^{b_2} \dots x_n^{b_n}} \\ \notag
&+ (-1)^{n}\sum_{b_1, b_2 \dots b_n = 1}^\infty
\frac{\la \Tr(M_2^r M_1^{b_1}) \Tr(M_1^{b_2}) \dots \Tr(M_1^{b_n}) \ra}{b_1 \dots b_n\
x_1^{b_1} \dots x_n^{b_n}},\ \ r > 0 \\ \notag
&Z^0_n (x_1, \dots, x_n):= 
(-1)^n\sum_{b_1, b_2 \dots b_n = 1}^\infty
\frac{\la \Tr(M_1^{b_1}) \dots \Tr(M_1^{b_n}) \ra}{b_1 \dots b_n\
x_1^{b_1} \dots x_n^{b_n}}
\end{align}
and the almost-fully principally-specialized wave function
 of level $r$ is
\begin{align}
Z^r(y, x) = \sum_{n = 0}^\infty \frac{1}{n!} Z^r_n(y, x, \dots x)
\end{align}
}

Note that with these definitions
\begin{equation}
Z^{(a)}(x) = x^{1/\hbar} Z^0 (x, x)
\end{equation}

\subsection{Loop equations in terms of $Z^r_n$}

Considering the coefficient in front of particular powers of
$1/x$ and $1/x_i$'s in loop equations \eqref{eq:loop1} and \eqref{eq:loop2}
we get the following equations relating particular formal matrix model correlators

\begin{align}
\label{eq:ward-identities-2MM}
&\la \Tr(M_1^{b_1}) \dots \Tr(M_1^{b_n}) \ra =\\ \notag
&\sum_{i=1}^a t_i \la \Tr\lb M_1^{b_1 - 1} M_2^{i-1}\rb \Tr(M_1^{b_2})\dots \Tr(M_1^{b_n}) \ra \\ \notag
&\la \Tr\lb M_2^{r} M_1^{b_1} \rb \Tr(M_1^{b_2})\dots \Tr(M_1^{b_n}) \ra =
\\ \notag
&\hbar \sum_{j = 2}^n b_j \la \Tr \lb M_2^{r - 1} M_1^{b_1 + b_j - 1} \rb
\Tr(M_1^{b_2})\dots \widehat{\Tr(M_1^{b_j})} \dots \Tr(M_1^{b_n})
 \ra \\ \notag
&
+ \hbar \sum_{p + q = b_1 - 1} \la \Tr(M_2^{r-1} M_1^p) \Tr (M_1^q)
\Tr(M_1^{b_2})\dots \Tr(M_1^{b_n}) \ra,
\end{align}
Here the hat above $\Tr(M_1^{b_j})$ means that it is excluded from the correlator.

Let us sum the above equations over all $b_1,\dots,b_n$ from $1$ to $\infty$ with the coefficient
$$
%\sum_{b_1 \dots b_n = 1}^\infty 
\frac{(-1)^n}{x_1^{b_1} b_2 \dots b_n
x_2^{b_2} \dots x_n^{b_n}}.$$
(note the absence of the $1/b_1$ factor).
%lead to the following system of PDE's for wave-functions $Z^r_n$:
We get:
\bl Loop equations, written in terms of $Z^r_n$, read
\begin{align}
\label{eq:set-of-loop-eqns}
&(-x_1 \frac{\d}{\d x_1}) Z^0_n(x_1, \dots , x_n) = \\ \notag
&\sum_{i = 1}^a t_i (-\frac{\d}{\d x_1}) Z^{i-1}_n(x_1, \dots ,x_n) -\dfrac{1}{\hbar x_1}Z^0_{n-1}(x_2,\dots,x_n),
\end{align}
\begin{align}
%\\ \notag
&\frac{1}{\hbar} (-x_1 \frac{\d}{\d x_1}) Z^1_n(x_1, \dots , x_n) = \\ \notag
&-\sum_{j=2}^n \left[ (-\frac{\d}{\d x_j}) Z^{0}_{n-1} (x_j, x_2, \dots \widehat{x_j} \dots x_n)   + \frac{1}{\hbar x_j} Z^0_{n-2}(x_2, \dots \widehat{x_j} \dots x_n)\right] \\ \notag
&+ \frac{2}{\hbar} (-\frac{\d}{\d x_1}) Z^{0}_n (x_1, \dots, x_n) - \frac{1}{\hbar^2 x_1} Z^0_{n-1}(x_2, \dots ,x_n) \\ \notag
&- x_1 \frac{\p^2}{\p u_1 \p u_2} \Big{|}_{u_1=u_2=x_1} Z^{0}_{n+1}(u_1, u_2, x_2, \dots, x_n) \\ \notag
& - \sum_{j=2}^n \frac{1}{(x_1 - x_j)} \left[ x_1 \frac{\d}{\d x_1}
Z^{0}_{n-1}(x_1, \dots \widehat{x_j} \dots x_n)
 - x_j \frac{\d}{\d x_j} Z^{0}_{n-1}(x_j, x_2, \dots \widehat{x_j} \dots x_n)\right],
\end{align}
and, for all $r > 1$,
\begin{align} \label{eq:setofloopeqns3}
%\\ \notag
&\frac{1}{\hbar} (-x_1 \frac{\d}{\d x_1}) Z^r_n(x_1, \dots , x_n) =\\ \notag
& -\sum_{j=2}^n (-\frac{\d}{\d x_j})
 Z^{r-1}_{n-1} (x_j, x_2, \dots \widehat{x_j} \dots x_n)
+ \frac{1}{\hbar} (-\frac{\d}{\d x_1}) Z^{r-1}_n (x_1, \dots, x_n) \\ \notag
& - x_1 \frac{\p^2}{\p u_1 \p u_2} \Big{|}_{u_1=u_2=x_1} Z^{r-1}_{n+1}(u_1, u_2, x_2, \dots, x_n) \\ \notag
& - \sum_{j=2}^n \frac{1}{(x_1 - x_j)} \left[ x_1 \frac{\d}{\d x_1}
 Z^{r-1}_{n-1}(x_1, x_2, \dots \widehat{x_j} \dots x_n)
- x_j \frac{\d}{\d x_j} Z^{r-1}_{n-1}(x_j, x_2, \dots \widehat{x_j} \dots x_n) \right].
\end{align}
\el

\subsection{Symmetrization of loop equations}
Last step to obtain quantum curve equation is to put all equations \eqref{eq:set-of-loop-eqns}--\eqref{eq:setofloopeqns3}
into principal specialization: put all $x_i$'s equal to $x$.

The following obvious statement plays a crucial role in the induction:

\bl
\label{lem:part-sym-funct}
Let $f(x_1|x_2,\dots,x_n)$ be a symmetric function in the variables $x_2,\dots, x_n$ (so, $x_1$ is treated specially here). Then we have the following formula for the derivative in the principal specialization.
\begin{align}
\frac{\p}{\p x} f(x|x, \dots, x) = \frac{\p}{\p u} \Big{|}_{u=x}f(u|x, \dots, x) + (n - 1) \frac{\p}{\p u} \Big{|}_{u=x} f(x|u, x, \dots, x).
\end{align}

In particular, if $f(x_1, x_2,\dots, x_n) = \frac{\p}{\p x_1} g(x_1, x_2,\dots,x_n)$, then
\begin{align}
& \frac{\p}{\p x} \frac{\p}{\p y} \Big{|}_{y = x} g(y| x,\dots, x)= 
\\ \notag & 
\frac{\p^2}{\p u^2} \Big{|}_{u=x} g(u|x,\dots,x)  + (n - 1) \frac{\p^2}{\p u_1 \p u_2} \Big{|}_{u_1=x,u_2=x}g(u_1|u_2,x,\dots,x).
\end{align}
\el

Since $Z^0_n$ is symmetric in all its arguments, the first equation of \eqref{eq:set-of-loop-eqns}
is equivalent to
\begin{align}
&\frac{1}{n} (- x \frac{\d}{\p x}) Z^0_n(x,\dots,x) = \\ \notag
&-t_1 \dfrac{1}{\hbar x} Z^0_{n-1}(x,\dots,x)
- \frac{1}{n}\frac{\d}{\p x} Z^0_n(x,\dots, x) \\ \notag
&+\sum_{i=2}^a t_i (-\frac{\d}{\p y})\Big{|}_{y=x} Z^{i-1}_n(y, x,\dots,x)
\end{align}
We multiply this by $\frac{1}{(n-1)!}$ and take the sum over $n \geq 0$. We have:
\begin{align}
\label{eq:pre-quantum-curve}
&(- x \frac{\d}{\p x}) Z^0 (x,\dots, x)=\\ \notag
&-t_1\left(\frac{\d}{\p x}+\dfrac{1}{\hbar x}\right)Z^0(x,\dots,x)\\ \notag
&+\sum_{i=2}^a t_i \sum_{n=0}^\infty \frac{1}{(n-1)!}(-\frac{\d}{\p y})\Big{|}_{y=x} Z^{i-1}_n(y, x,\dots,x).
\end{align}
Then, the existence of a quantum curve equation relies on two observations:

\bl \label{lem:quantum-curve-recursion}
 We have:
\begin{align}
i > 1:\  & \sum_{n=0}^\infty \frac{1}{(n-1)!}(-\frac{\d}{\p y})\Big{|}_{y=x} Z^{i}_n(y, x,\dots,x) \\ \notag & = 
\lb \frac{1}{x} + \hbar \frac{\d}{\p x} \rb \sum_{n=0}^\infty \frac{1}{(n-1)!}(-\frac{\d}{\p y})\Big{|}_{y=x} Z^{i-1}_n(y, x,\dots,x)
\\ \notag
i = 1:\  & \sum_{n=0}^\infty \frac{1}{(n-1)!}(-\frac{\d}{\p y})\Big{|}_{y=x} Z^{i}_n(y, x,\dots,x) \\ \notag & =
\hbar\lsb -\frac{\d^2}{\p x^2} - \frac{2}{\hbar x} \frac{\d}{\p x} - \frac{1/\hbar (1/\hbar -1)}{x^2} \rsb Z^0 
\\ \notag &
= -\frac{1}{\hbar}\lb \frac{1}{x} + \hbar \frac{\d}{\p x} \rb^2 Z^0
\end{align}
\el
\begin{proof}
These equations are direct corollaries of Equations~\eqref{eq:set-of-loop-eqns}, we just have to 
put them into principal specialization and apply
Lemma \ref{lem:part-sym-funct}.
\end{proof}

We combine Equation~\eqref{eq:pre-quantum-curve} and Lemma~\eqref{lem:quantum-curve-recursion}, and we obtain the following equation:
\begin{align}
(-\hbar x \frac{\d}{\p x}) Z^0 = -\sum_{i=1}^a t_i \lb \frac{1}{x} + \hbar \frac{\d}{\p x} \rb^i Z^0.
\end{align}
which, with help of commutation relation
\begin{align}
x^{1/\hbar} \lb \frac{1}{x} + \hbar \frac{\d}{\p x} \rb = \hbar \frac{\d}{\p x}\circ  x^{1/\hbar},
\end{align}
leads directly to the statement of Theorem~\ref{thm:quantum-curve}.

\section{4-colored maps and 4-matrix models} \label{section:4case}

It turns out that the ideas above can be applied not only to bi-colored maps (which correspond to the 2-matrix model case), but also to 4-colored maps. In the current section we outline the idea of the proof of the spectral curve topological recursion for the enumeration of 4-colored maps. 

4-colored maps arise as a natural generalization of bi-colored maps. Instead of considering partitions of surfaces into black and white polygons, we consider partitions into polygons of four colors $c_1,c_2,c_3,c_4$, such that polygons of color $c_1$ are glued only to polygons of color $c_2$, polygons of color $c_2$ are glued only to polygons of colors $c_1$ and $c_3$, polygons of color $c_3$ are only glued to those of color $c_2$ and $c_4$ and finally polygons of color $c_4$ are only glued to polygons of color $c_3$. This can be represented in terms of the following color incidency matrix:
\begin{equation}
\left(\begin{array}{cccc}
0 & 1 & 0 & 0 \\
1 & 0 & 1 & 0 \\
0 & 1 & 0 & 1 \\
0 & 0 & 1 & 0 
\end{array}\right)
\end{equation}

Applying considerations similar to the ones in the above sections, it's easy to see that the problem of enumeration of such 4-colored maps is governed by a 4-matrix model with the interaction part of the potential being equal to
\begin{equation}
-N\Tr(M_1 M_2 - M_1 M_4 + M_3 M_4),
\end{equation}
since the inverse  of the above incidency matrix is equal to
\begin{equation}
\left(
\begin{array}{cccc}
 0 & 1 & 0 & -1 \\
 1 & 0 & 0 & 0 \\
 0 & 0 & 0 & 1 \\
 -1 & 0 & 1 & 0 \\
\end{array}
\right)
\end{equation}
We see that in the 4-colored maps case, after a renumeration of matrices and a certain change of signs, this still gives us the matrix model for a chain of matrices (which is no longer true for, e.g., 6-colored maps). Fortunately, the case of matrix model for a chain of matrices was studied by Eynard in \cite{EynardChain}, and the master loop equation obtained there gives rise to the spectral curve topological recursion for this problem.
Again, it's easy to see in the analogous way to what was discussed in the previous sections that the individual building blocks of loop equations can be proved to hold by purely combinatorial means.

\section*{Acknowledgements}
N.~O. wants to thank G.~Borot and B.~Eynard for numerous discussions on this subject. In particular, we would like to mention that most of the results of this paper were derived independently by Borot~\cite{Borotprivate} while this paper was being written, and the combinatorial approach to loop equations in Section~\ref{section:loopeqs} was also independently derived by Eynard~\cite[Chapter 8]{Eynbook}. We would also like to thank B.~Eynard for pointing towards some extra missing references in a preliminary version.

For N~.O., it is a pleasure to thank the Korteweg-de~Vries Institute for Mathematics for its warm welcome at an early stage of this work.

P.~D.-B., A.~P., and S.~S. were supported by the Netherlands Organization for Scientific Research (NWO). P.~D.-B. and A.~P. were also partially supported by the Russian President's Grant of Support for the Scientific Schools NSh-3349.2012.2 and by RFBR grant 15-31-20832-mol\_a\_ved; P.~D.-B. was partially supported by RFBR grants 13-02-00478, 14-01-31395-mol\_a and 15-01-05990, RFBR-India grant 14-01-92691-Ind\_a and RFBR-Turkey grant 13-02-91371-St\_a; A.~P. was partially supported by RFBR grants 13-02-00457 and 14-01-31492-mol\_a. P.~D.-B. was supported by an MPIM Bonn fellowship.

\end{document}